\documentclass{PoS}
\usepackage{amsmath}
\usepackage{cancel}
\usepackage{verbatim}   

\newcommand{\al}{\ensuremath{\alpha} }
\newcommand{\be}{\ensuremath{\beta} }
\newcommand{\ga}{\ensuremath{\gamma} }
\newcommand{\gaEff}{\ensuremath{\ga_{\rm eff}} }
\newcommand{\De}{\ensuremath{\Delta} }
\newcommand{\eps}{\ensuremath{\epsilon} }
\newcommand{\la}{\ensuremath{\lambda} }
\newcommand{\Si}{\ensuremath{\Sigma} }
\newcommand{\om}{\ensuremath{\omega} }
\newcommand{\X}{\ensuremath{\!\times\!} }
\newcommand{\gsim}{\ensuremath{\gtrsim} }
\newcommand{\lsim}{\ensuremath{\lesssim} }
\newcommand{\vev}[1]{\ensuremath{\left\langle #1 \right\rangle} }
\newcommand{\psibar}{\ensuremath{\overline\psi} }

\newcommand{\pbpv}{\ensuremath{\vev{\psibar_v\psi_v}} }
\newcommand{\Sb}{\ensuremath{\cancel{S^4}} }
\newcommand{\eq}[1]{Eq.~\ref{#1}}
\newcommand{\fig}[1]{Fig.~\ref{#1}}
\newcommand{\refcite}[1]{Ref.~\cite{#1}}

\newcommand{\mysection}[1]{\section{#1}}
\newcommand{\mutesection}[1]{\section*{#1}}

\title{Determining the mass anomalous dimension through the eigenmodes of Dirac operator}
\ShortTitle{Mass anomalous dimension from Dirac eigenmode scaling}

\author{\speaker{Anqi Cheng}, Anna Hasenfratz, Gregory Petropoulos, David Schaich\footnote{Present address: Department of Physics, Syracuse University, Syracuse, NY 13244} \\
        Department of Physics, University of Colorado, Boulder, CO 80309 \\
        E-mail: \email{anqi.cheng@colorado.edu}}

\abstract{We define a scale-dependent effective mass anomalous dimension from the scaling of the mode number of the massless Dirac operator, which connects the perturbative $\ga_m$ of an asymptotically-free system to the universal $\ga_m^{\star}$ at a conformal fixed point.  We use a stochastic algorithm to measure the mode number up to the cutoff scale on lattices as large as $48^4$.  Focusing on SU(3) lattice gauge theory with $N_f = 12$ massless fundamental fermions, we examine systematic effects due to finite volumes and non-zero fermion masses.  Our results suggest the existence of an infrared fixed point with $\ga_m^{\star} \approx 0.25$.  Our method provides a unique probe to study systems from the UV to the IR.  It is universal and can be applied to any lattice model of interest, including both chirally-broken and IR-conformal systems.}

\FullConference{31st International Symposium on Lattice Field Theory - LATTICE 2013 \\
                29 July -- 3 August 2013 \\
                Mainz, Germany}

\begin{document}
Near-conformal gauge--fermion systems have received significant attention in recent years.
Some of these models remain phenomenologically viable candidates for new physics beyond the standard model, while others are simply interesting non-perturbative quantum field theories.
The infrared dynamics of most of these models is strongly coupled, motivating lattice studies -- for a limited set of references see Refs.~\cite{Appelquist:2009ty, Deuzeman:2009mh, Fodor:2011tu, Appelquist:2011dp, Hasenfratz:2011xn, DeGrand:2011cu, Cheng:2011ic, Fodor:2012uw, Fodor:2012et, Aoki:2012eq, Itou:2012qn, Lin:2012iw, Jin:2012dw, Cheng:2013eu, Aoki:2013xza, Hasenfratz:2013uha, Aoki:2013qxa, Hasenfratz:2013eka}. 

In principle lattice calculations can distinguish the chirally-broken and IR-conformal scenarios.
However, lattice methods that are effective to study QCD-like systems frequently are not optimal to investigate near-conformal dynamics.
The available lattice volumes limit the accessible energy range, and chirally-broken systems with a slowly-running gauge coupling of the sort expected near the conformal window can appear very similar to infrared conformal dynamics.
The most promising methods explore differences between the two scenarios at strong gauge coupling.
Backward flows in running coupling or Monte Carlo renormalization group calculations are examples of such distinctive signals.
In this paper we explore another promising approach, the scale dependence of an effective mass anomalous dimension \gaEff extracted from the mode number of the massless Dirac operator.
This quantity allows us to probe the systems from the UV to the IR, making it possible to distinguish IR-conformal dynamics from chirally-broken models near the conformal window.

Some results from this approach have been published in \refcite{Cheng:2013eu}.
Here we expand our theoretical discussion and show new results obtained using a stochastic algorithm to measure the mode number~\cite{Giusti:2008vb}, which we recently implemented for staggered fermions.
We concentrate on the SU(3) 12-flavor system, though the method we use is applicable to any other system and \refcite{Cheng:2013eu} contains results for $N_f = 4$ and 8 as well.

\mysection{Scaling of the mode number} 
The spectral density of the Dirac operator is related to the partially quenched chiral condensate
\begin{equation}
  \rho(\la) = \frac{1}{\pi} \lim_{\eps \to 0} \mbox{Re}\int {d^4 x}\pbpv|_{m_v = i\la + \eps},
\end{equation}
where \pbpv is evaluated at the imaginary valence quark mass $m_v = i\la + \eps$.
In a chirally-broken system this leads to the Banks--Casher relation $\Si = \lim_{\la \to 0} \pi\rho(\la)$ in the infinite-volume chiral limit~\cite{Banks:1979yr}.
Conformal systems are chirally symmetric, $\rho(0) = 0$.
General critical scaling suggests
\begin{equation}
  \label{eq:rho_scaling}
  \rho(\la) \propto \la^{\al}
\end{equation}
in the small-$\la$ infrared limit.

In asymptotically-free systems, one-loop perturbation theory predicts
\begin{align}
  \label{eq:rho_PT}
  \rho(g^2) & = C \la^{A(g^2)} &
  A(g^2) & \equiv \frac{4}{1 + \ga_m(g^2)} - 1
\end{align}
in the ultraviolet, where
\begin{equation}
  \ga_m(g^2) = \frac{6C_2(R)}{16\pi^2} g^2 
\end{equation}
is the universal one-loop mass anomalous dimension for fermions in representation $R$.\newpage 

The integral of the spectral density, the mode number $\nu$, is renormalization group (RG) invariant~\cite{Giusti:2008vb, DelDebbio:2010ze}.
In the infrared limit of a conformal system, \eq{eq:rho_scaling} predicts
\begin{equation}
  \label{eq:nu}
  \nu(\la) = V\int^{\la}_{-\la}\rho(\om)d\om \propto V \la^{\al + 1},
\end{equation}
relating the exponent \al and the anomalous dimension at the infrared fixed point (IRFP)~\cite{DelDebbio:2010ze, Patella:2012da},
\begin{equation}
  \label{eq:al_ga}
  \al + 1 = \frac{4}{1 + \ga_m^{\star}}.
\end{equation}
Comparing Eqs.~\ref{eq:rho_scaling}, \ref{eq:rho_PT} and \ref{eq:al_ga} suggests the extension of the scaling relation \eq{eq:nu} as
\begin{equation}
  \label{eq:nu_ext}
  \nu(\la) = C V \la^{\al(\la) + 1} = C V \la^{\frac{4}{1 + \gaEff(\la)}}.
\end{equation}
That is, we allow \al to depend on $\la$, and interpret $\gaEff(\la)$ as a scale-dependent effective mass anomalous dimension.
$\gaEff(\la)$ is anchored by the universal one-loop perturbative $\ga_m$ in the ultraviolet, and by $\ga_m^{\star}$ in the infrared limit.

\begin{figure}[tb]
  \includegraphics[width=0.45\linewidth]{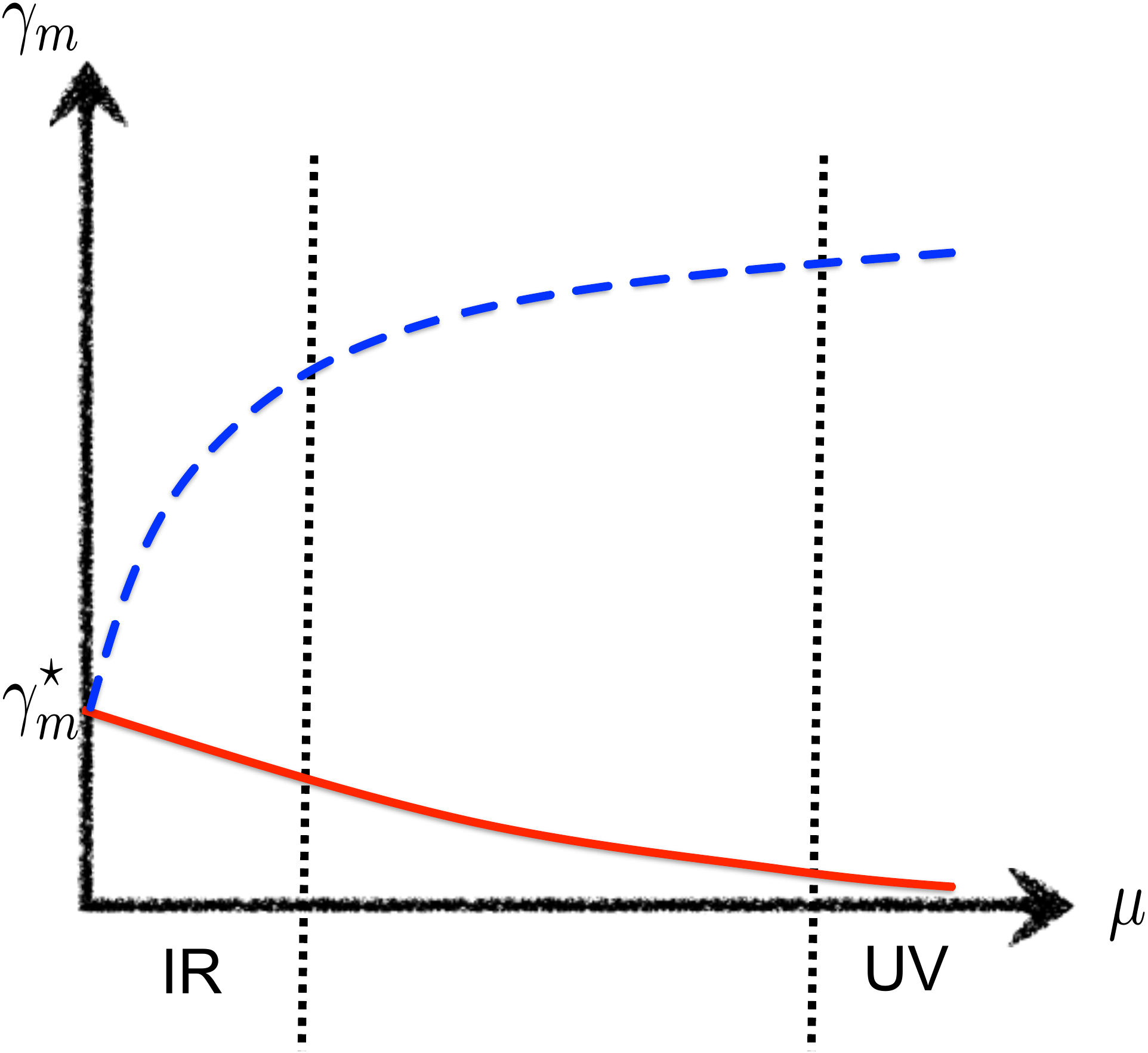}\hfill
  \includegraphics[width=0.45\linewidth]{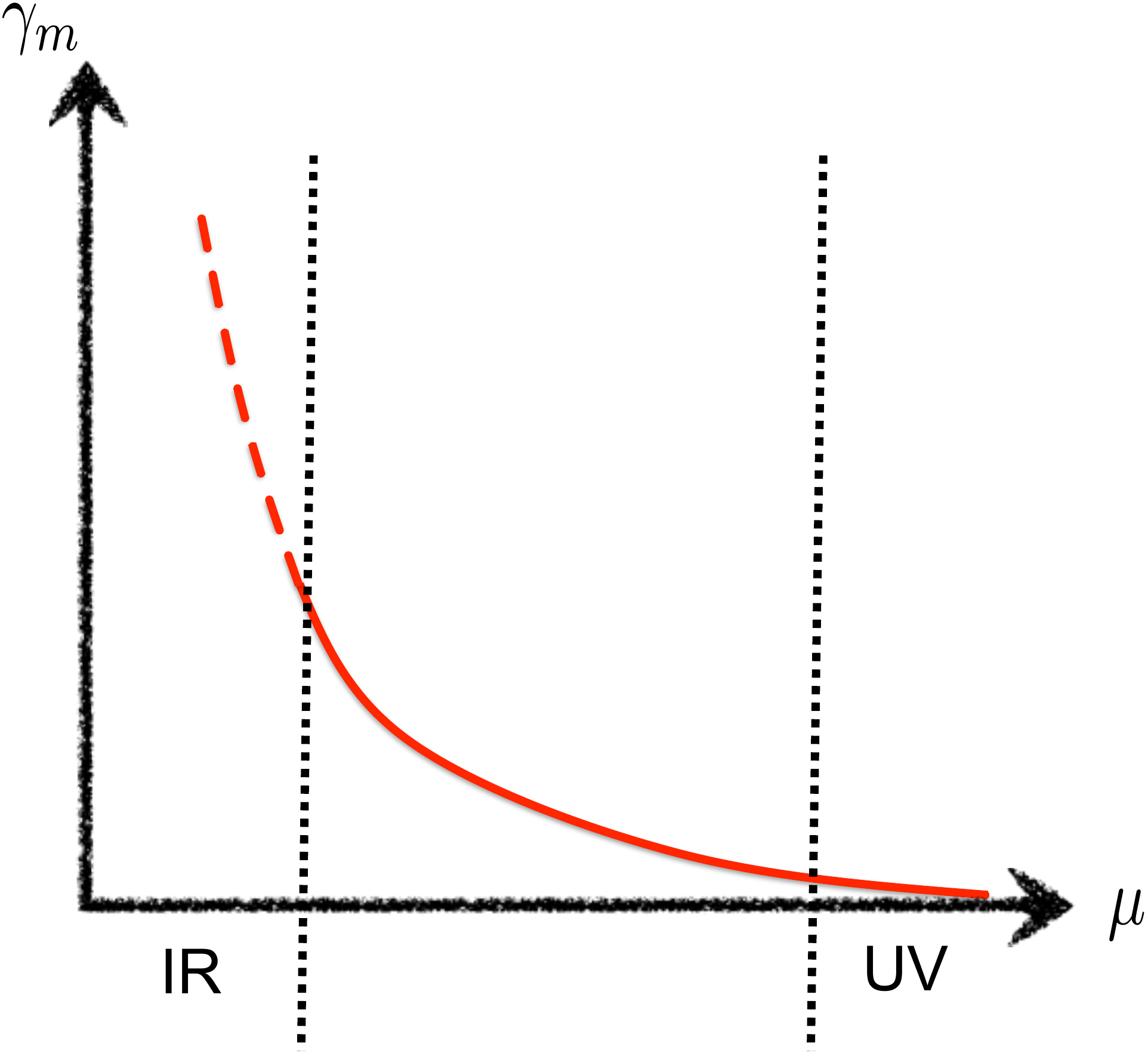}
  \caption{\label{fig:cartoon} Cartoons of the effective mass anomalous dimension $\gaEff(\la)$ in IR-conformal (left) and chirally-broken (right) continuum systems.  In both cases, \gaEff starts from zero at the asymptotically-free UV fixed point and increases as the energy decreases.  In IR-conformal systems, $\gaEff \to \ga_m^{\star}$ at the IRFP as $\la \to 0$.  At even stronger coupling \gaEff might increase with energy as suggested by the dashed blue line.  In chirally-broken systems the scaling form $\rho(\la) \propto \la^{\al(\la)}$ is lost below the chiral symmetry breaking scale.}
\end{figure}

\fig{fig:cartoon} sketches $\gaEff(\la)$ for idealized (infinite volume, zero-mass, continuum) IR-conformal and chirally-broken systems.
The solid red line of the left panel corresponds to the mass anomalous dimension of an IR-conformal system in between the perturbative and infrared fixed points.
At large \la we expect to see the ultraviolet behavior $\ga_m \to 0$, while in the $\la \to 0$ infrared limit $\gaEff \to \ga_m^{\star}$.
There is no perturbative guidance at strong coupling.
Since we expect a backward flow in the gauge coupling, it is conceivable that $\gaEff(\la)$ may increase towards the ultraviolet (blue dashed curve).
The right panel shows the equivalent behavior of a chirally-broken system.
The effective anomalous dimension has the same ultraviolet behavior, even though it loses its physical meaning below the energy scale of chiral symmetry breaking.
We denote this rather ill-defined situation by a dashed line.
The qualitative difference between the two panels of \fig{fig:cartoon} indicates how the mode number could distinguish conformal and chirally-broken systems.

\mysection{Computational details} 
The investigation of the Dirac eigenmodes is part of our ongoing study of the 4-, 8- and 12-flavor SU(3) systems~\cite{Hasenfratz:2011xn, Cheng:2011ic, Cheng:2013eu, Hasenfratz:2013uha, Hasenfratz:2013eka}.
We use nHYP-smeared staggered fermions and our gauge action is a combination of fundamental and adjoint plaquette terms with $\be_A / \be_F = -0.25$.
In this paper we concentrate on the 12-flavor system.

Our previous investigations of $N_f = 8$ and 12 revealed an unusual strong-coupling phase where the single-site shift symmetry of the staggered lattice action is spontaneously broken (\Sb phase)~\cite{Cheng:2011ic, Hasenfratz:2013uha}.
In the present study we stay on the weak-coupling side of this phase, though our strongest gauge coupling, $\be_F = 3.0$, is very close to it.
In \refcite{Cheng:2013eu} we used configurations with periodic boundary conditions in the spatial directions and a small but non-vanishing fermion mass $m = 0.0025$.
Our new results are obtained using $m = 0$ with anti-periodic boundary conditions in all four directions.
In order to monitor finite-volume effects we compare $24^4$ and $32^4$ volumes at every gauge coupling value.
At our weakest coupling $\be_F = 6.0$, where finite-volume effects are the most severe, we also use a $48^4$ ensemble.

We measure the mode number using the stochastic method developed in \refcite{Giusti:2008vb}.
Using 20--40 configurations from each ensemble, we evaluate the mode number with five stochastic sources at each value of $\la$.
With values of \la separated by $\De_\la = 0.005$, we cover the eigenvalue range up to $\la = 0.5$.\footnote{In this paper we adopt the conventional continuum normalization of the eigenvalue $\la$, which differs by a factor of two compared to the Dirac operator normalization used in the MILC code and in \refcite{Cheng:2013eu}.}
To extract the scale-dependent $\al(\la)$ and $\gaEff(\la)$ from the mode number, we simply perform a linear fit to the logarithms
\begin{equation}
  \label{eq:fit_form}
  \log\left[\nu(\la)\right] = (\al(\la) + 1)\log\left[\la\right] + \mbox{ constant},
\end{equation}
using finite intervals in \la and a jackknife analysis to determine uncertainties.
While the mode number $\nu$ can have larger finite-volume corrections than its $\la$-derivative, the spectral density $\rho$, we prefer to work with $\nu$.
Differentiating \eq{eq:nu_ext} to obtain $\rho$ introduces $\frac{d\al}{d\la}$ corrections that can lead to significant systematic effects.

\mysection{Finite-volume and finite-mass effects} 
\begin{figure}[tb]
  \includegraphics[width=0.45\linewidth]{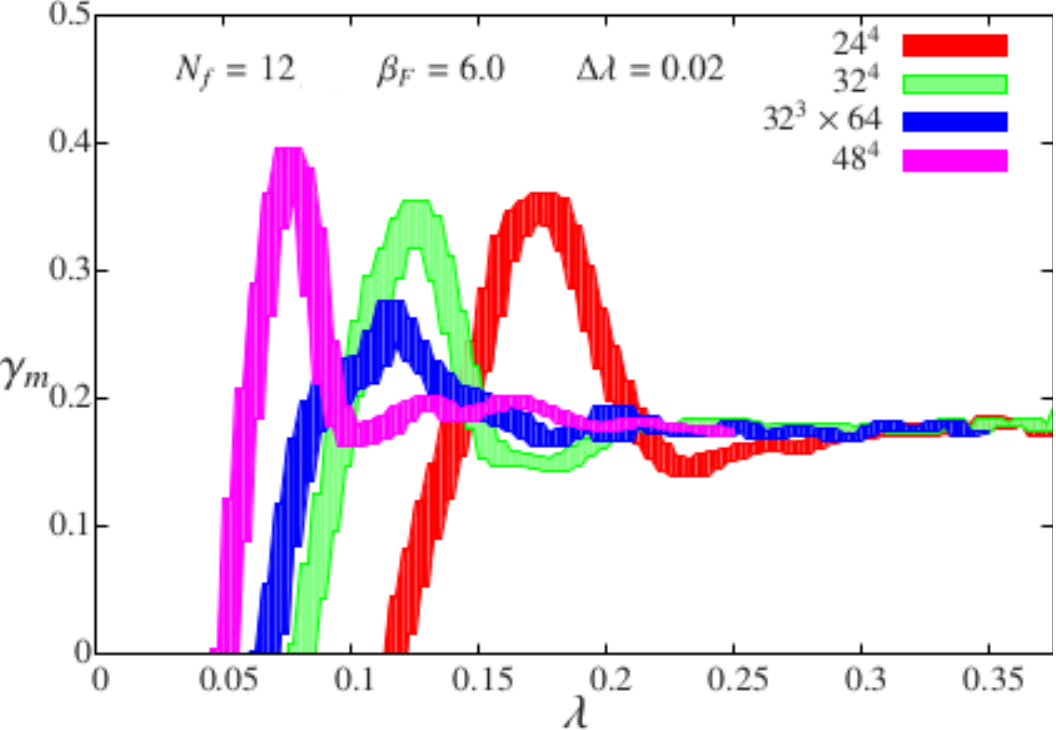}\hfill
  \includegraphics[width=0.45\linewidth]{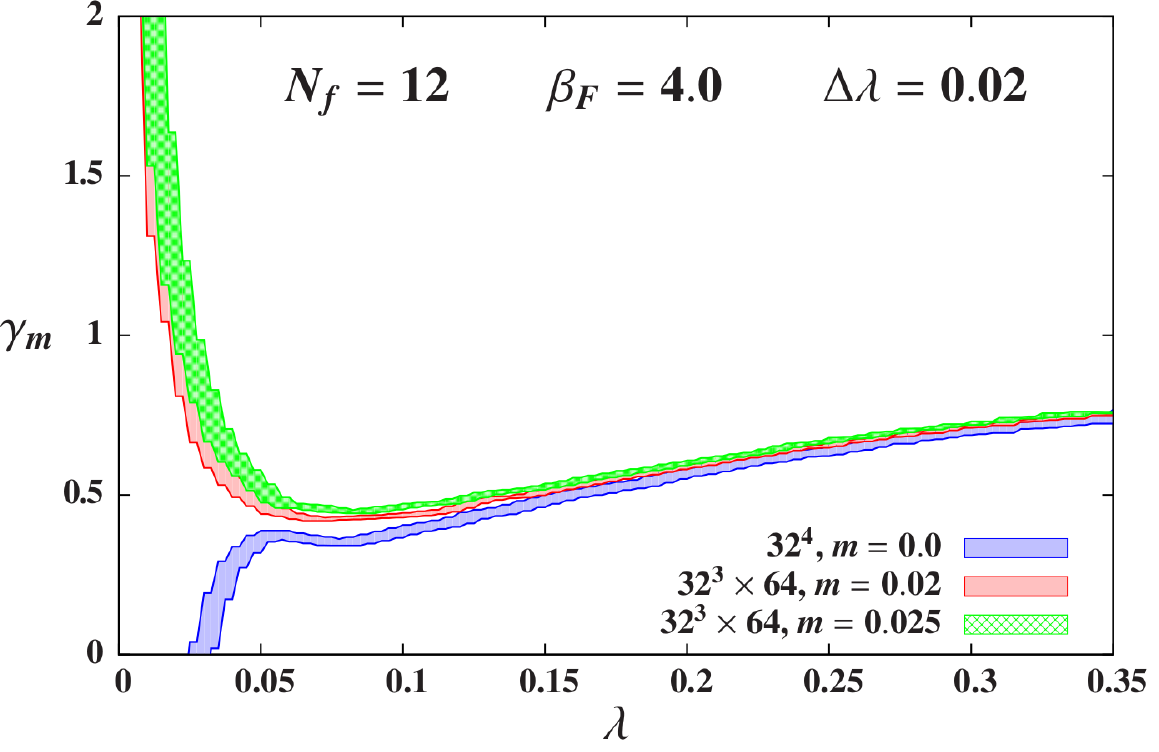}
  \caption{\label{fig:fs_test}Tests of finite-volume effects (left) and finite-mass effects (right).  Both plots show the effective mass anomalous dimension \gaEff as a function of the eigenvalue $\la$, extracted according to \protect\eq{eq:fit_form} and \protect\eq{eq:al_ga}.}
\end{figure}

It is important to monitor finite-volume corrections to \eq{eq:fit_form}.
The left panel of \fig{fig:fs_test} shows the predicted effective mass anomalous dimension at $\be_F = 6.0$ on $24^4$, $32^4$, $32^3\X64$ and $48^4$ volumes.
The $L^4$ ensembles use $m = 0$ with anti-periodic boundary conditions in all four directions, while the $32^3\X64$ ensemble uses periodic spatial boundary conditions with $m = 0.0025$.
At large enough $\la \gsim 0.3$ the predictions from the smallest $L = 24$ agree with the results from larger volumes.
For the largest $L = 48$, finite-volume effects appear manageable for $\la \gsim 0.15$ even at $\be_F = 6.0$, our weakest coupling with the most severe finite-volume effects.
\fig{fig:fs_test} also indicates that finite-volume effects on our new $L^4$ ensembles with anti-periodic boundary conditions are more significant than on the $L^3\X2L$ ensembles studied in \refcite{Cheng:2013eu}.

While we use $m = 0$ in our calculations, it is instructive to consider how finite mass affects the mode number and predicted $\gaEff$.
The right panel of \fig{fig:fs_test} considers two $32^3\X64$ ensembles with periodic spatial boundary conditions and $m = 0.02$ and 0.025.
At small \la the predicted anomalous dimension becomes unphysically large, $\gaEff \gsim 2$, as \eq{eq:rho_scaling} breaks down due to the explicit chiral symmetry breaking from these non-zero fermion masses.
Even so, as \la increases the results we obtain on these ensembles converge with the prediction from a $32^4$ ensemble with $m = 0$, all of which show \gaEff increasing with $\la$.

\mysection{Results with $N_f=12$ fundamental fermions} 
In \refcite{Cheng:2013eu} we showed that for the 4-flavor SU(3) theory our definition of the effective mass anomalous dimension \gaEff shows continuum-like scaling.
$\gaEff(\la)$ qualitatively follows the sketch in the right panel of \fig{fig:cartoon}, and is consistent with the one-loop perturbative prediction for $\la \gsim 0.2$ (corresponding to $\ga_m(g^2) \lsim 0.4$).

\begin{figure}[tb]
  \includegraphics[width=0.45\linewidth]{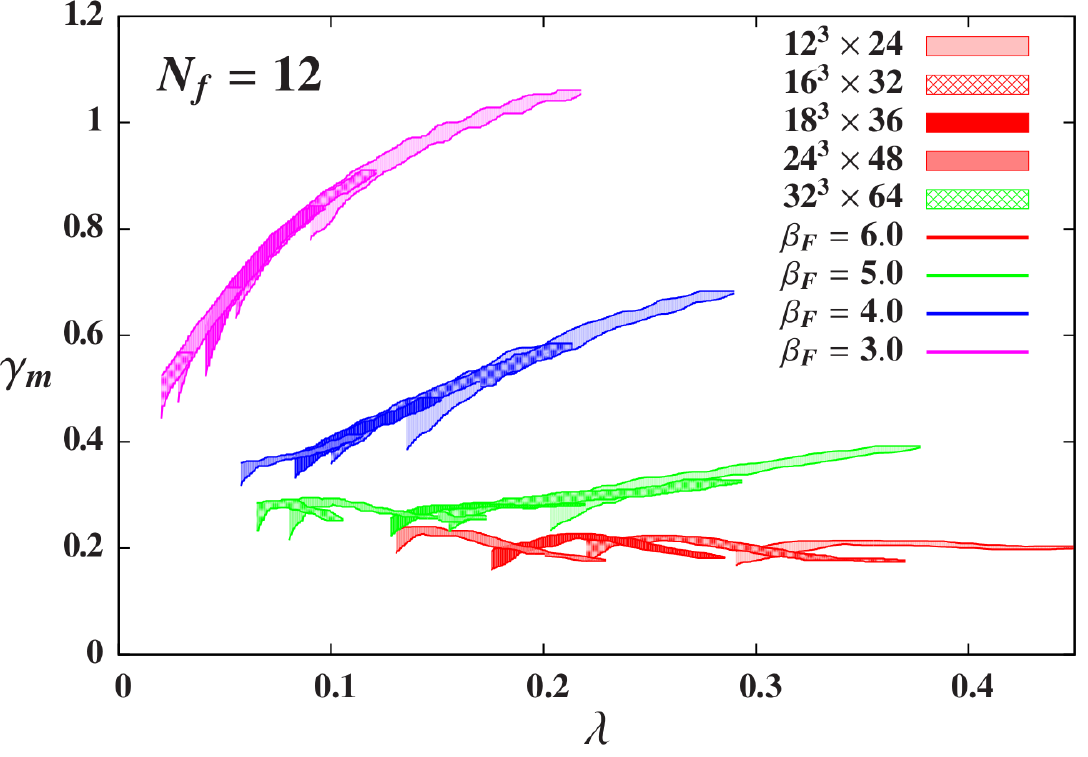}\hfill
  \includegraphics[width=0.45\linewidth]{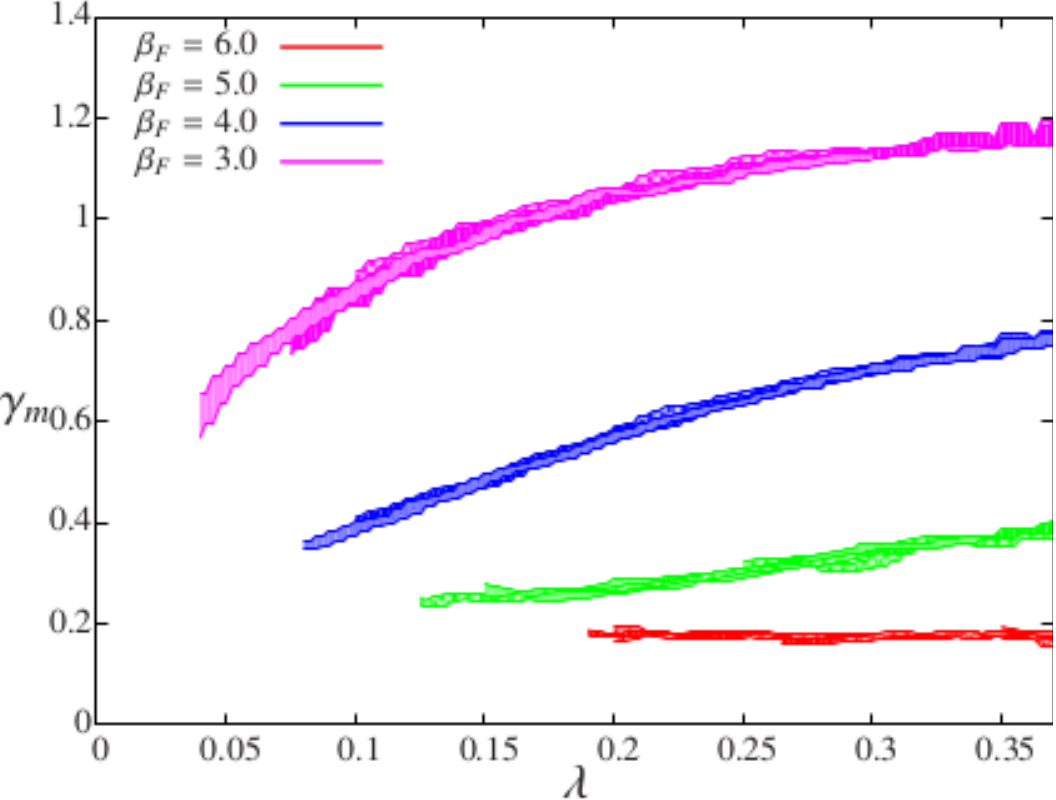}
  \caption{\label{fig:12f}The effective mass anomalous dimension \gaEff for $N_f = 12$, determined either from direct eigenvalue calculations (left, from \protect\refcite{Cheng:2013eu}) or using stochastic mode number measurements (right).  We carry out stochastic measurements on lattice volumes up to $32^4$ and (for $\be_F = 6.0$) $48^4$.}
\end{figure}

\fig{fig:12f} summarizes our results for the effective anomalous dimension with 12 fundamental flavors.
The left panel is from \refcite{Cheng:2013eu} using direct eigenvalue calculations on several different volumes, while the right panel shows our new results obtained using the stochastic mode number calculation.
In both plots we only present results that appear largely volume independent.
We consider gauge coupling values $3.0 \leq \be_F \leq 6.0$, spanning a significant range from near the \Sb phase to the weakest coupling where finite-volume effects remain manageable on $32^4$ and $48^4$ lattices.

The results at strong gauge coupling are striking.
They do not follow the behavior expected near the asymptotically-free fixed point: \gaEff increases towards the ultraviolet for all $\be_F \leq 5.0$.
This behavior is not consistent with ultraviolet dynamics that is driven by the perturbative fixed point, and implies that all couplings $\be_F \leq 5.0$ are outside the scaling regime of this UVFP.

Numerical calculations cannot exclude the possibility that some unexpected and unusual property of a chirally-broken system produces the behavior of \gaEff depicted in \fig{fig:12f}.
However, our results are clearly consistent with the existence of a conformal IR fixed point.
They follow the conjectured behavior in the left panel of \fig{fig:cartoon}, which we motivated from a backward flow in the gauge coupling.
The energy dependence of the effective mass anomalous dimension is minimal at $\be_F = 6.0$, suggesting that this gauge coupling is close to the fixed point $\be_F^{\star}$ in the scheme defined by this observable.
All of our results for $3.0 \leq \be_F \leq 6.0$ can be consistently extrapolated to the $\la = 0$ infrared limit to predict the scheme-independent mass anomalous dimension $\ga_m^{\star}$ of the IRFP.
We have not yet completed a careful $\la \to 0$ extrapolation of our new results, but our preliminary analysis indicates a relatively small value $\ga_m^{\star} \approx 0.25$.

\section{Conclusion} 
The RG invariance of the Dirac eigenmode number allows the determination of a scale-dependent effective mass anomalous dimension that connects the perturbative $\ga_m(g^2)$ to the universal $\ga_m^{\star}$ of a conformal system in the infrared limit.
In the ultraviolet the mode number scales with the perturbative anomalous dimension $\ga_m(g^2)$ if the UV dynamics is driven by the asymptotically-free perturbative fixed point.
This behavior is expected both in the scaling regime of chirally-broken systems as well as in IR-conformal systems at weak coupling.
On the other hand, conformal systems at strong coupling that exhibit backward flow could show very different behavior, offering a promising method to distinguish conformal and chirally-broken systems.

Our investigations of the $N_f = 4$ system in \refcite{Cheng:2013eu} predicted scaling consistent with the perturbative fixed point.
We observe very different behavior with 12 flavors.
At strong gauge coupling \gaEff increases towards the ultraviolet, and both the strong- and weak-coupling data are consistent with the existence of a conformal fixed point in the $\la \to 0$ infrared limit.
While we work with exactly massless fermions and monitor finite-volume effects, the $\la \to 0$ IR extrapolation of our results is non-trivial and remains under investigation.
It is clear from \fig{fig:12f} that studying multiple gauge couplings is crucial for the accurate determination of $\ga_m^{\star}$ at $\la = 0$.
Some related issues were recently discussed by \refcite{Landa-Marban:2013oia} in the context of the two-flavor Schwinger model.

\mutesection{Acknowledgments} 
We thank Agostino Patella for advice on developing code to stochastically calculate the mode number, as well as K.~Splittorff, C.~Lehner and T.~DeGrand for discussions regarding the perturbative evaluation of the spectral density.
A.~H.\ is grateful for the hospitality of the Brookhaven National Laboratory HET group during her extended visit.
This research was partially supported by the U.S.~Department of Energy (DOE) through Grant No.~DE-SC0010005 and by the DOE Office of Science Graduate Fellowship Program under Contract No.~DE-AC05-06OR23100 (G.~P.).
Our code is based in part on the MILC Collaboration's public lattice gauge theory software.\footnote{\href{http://www.physics.utah.edu/~detar/milc/}{http://www.physics.utah.edu/$\sim$detar/milc/}}
Numerical calculations were carried out on the HEP-TH and Janus clusters at the University of Colorado, the latter supported by the U.S.~National Science Foundation (NSF) through Grant No.~CNS-0821794; at Fermilab through USQCD supported by the DOE; and at the San Diego Computing Center and Texas Advanced Computing Center through XSEDE supported by NSF Grant No.~OCI-1053575.

{\renewcommand{\baselinestretch}{0.86} 
  \bibliographystyle{utphys}
  \bibliography{modenumber}

\providecommand{\href}[2]{#2}\begingroup\raggedright\begin{thebibliography}{10}

\bibitem{Appelquist:2009ty}
T.~Appelquist, G.~T. Fleming, and E.~T. Neil
  \href{http://dx.doi.org/10.1103/PhysRevD.79.076010}{{\em Phys. Rev.} {\bf
  D79} (2009)  076010}, \href{http://arxiv.org/abs/0901.3766}{{\tt
  arXiv:0901.3766}}.

\bibitem{Deuzeman:2009mh}
A.~Deuzeman, M.~P. Lombardo, and E.~Pallante
  \href{http://dx.doi.org/10.1103/PhysRevD.82.074503}{{\em Phys. Rev.} {\bf
  D82} (2010)  074503}, \href{http://arxiv.org/abs/0904.4662}{{\tt
  arXiv:0904.4662}}.

\bibitem{Fodor:2011tu}
Z.~Fodor, K.~Holland, J.~Kuti, D.~Nogradi, and C.~Schroeder
  \href{http://dx.doi.org/10.1016/j.physletb.2011.07.037}{{\em Phys. Lett.}
  {\bf B703} (2011)  348--358}, \href{http://arxiv.org/abs/1104.3124}{{\tt
  arXiv:1104.3124}}.

\bibitem{Appelquist:2011dp}
T.~Appelquist, G.~T. Fleming, M.~F. Lin, E.~T. Neil, and D.~Schaich
  \href{http://dx.doi.org/10.1103/PhysRevD.84.054501}{{\em Phys. Rev.} {\bf
  D84} (2011)  054501}, \href{http://arxiv.org/abs/1106.2148}{{\tt
  arXiv:1106.2148}}.

\bibitem{Hasenfratz:2011xn}
A.~Hasenfratz \href{http://dx.doi.org/10.1103/PhysRevLett.108.061601}{{\em
  Phys. Rev. Lett.} {\bf 108} (2012)  061601},
  \href{http://arxiv.org/abs/1106.5293}{{\tt arXiv:1106.5293}}.

\bibitem{DeGrand:2011cu}
T.~DeGrand \href{http://dx.doi.org/10.1103/PhysRevD.84.116901}{{\em Phys. Rev.}
  {\bf D84} (2011)  116901}, \href{http://arxiv.org/abs/1109.1237}{{\tt
  arXiv:1109.1237}}.

\bibitem{Cheng:2011ic}
A.~Cheng, A.~Hasenfratz, and D.~Schaich
  \href{http://dx.doi.org/10.1103/PhysRevD.85.094509}{{\em Phys. Rev.} {\bf
  D85} (2012)  094509}, \href{http://arxiv.org/abs/1111.2317}{{\tt
  arXiv:1111.2317}}.

\bibitem{Fodor:2012uw}
Z.~Fodor, K.~Holland, J.~Kuti, D.~Nogradi, C.~Schroeder, and C.~H. Wong
  \href{http://pos.sissa.it/archive/conferences/164/025/Lattice
  2012_025.pdf}{{\em PoS} {\bf Lattice 2012} (2012)  025},
  \href{http://arxiv.org/abs/1211.3548}{{\tt arXiv:1211.3548}}.

\bibitem{Fodor:2012et}
Z.~Fodor, K.~Holland, J.~Kuti, D.~Nogradi, C.~Schroeder, and C.~H. Wong
  \href{http://pos.sissa.it/archive/conferences/164/279/Lattice
  2012_279.pdf}{{\em PoS} {\bf Lattice 2012} (2012)  279},
  \href{http://arxiv.org/abs/1211.4238}{{\tt arXiv:1211.4238}}.

\bibitem{Aoki:2012eq}
Y.~Aoki, T.~Aoyama, M.~Kurachi, T.~Maskawa, K.-i. Nagai, H.~Ohki, A.~Shibata,
  K.~Yamawaki, and T.~Yamazaki
  \href{http://dx.doi.org/10.1103/PhysRevD.86.054506}{{\em Phys. Rev.} {\bf
  D86} (2012)  054506}, \href{http://arxiv.org/abs/1207.3060}{{\tt 1207.3060}}.

\bibitem{Itou:2012qn}
E.~Itou \href{http://dx.doi.org/10.1093/ptep/ptt053}{{\em PTEP} {\bf 2013}
  (2013)  083B01}, \href{http://arxiv.org/abs/1212.1353}{{\tt
  arXiv:1212.1353}}.

\bibitem{Lin:2012iw}
C.-J.~D. Lin, K.~Ogawa, H.~Ohki, and E.~Shintani
  \href{http://dx.doi.org/10.1007/JHEP08(2012)096}{{\em JHEP} {\bf 1208} (2012)
   096}, \href{http://arxiv.org/abs/1205.6076}{{\tt arXiv:1205.6076}}.

\bibitem{Jin:2012dw}
X.-Y. Jin and R.~D. Mawhinney {\em PoS} {\bf Lattice 2011} (2012)  066,
  \href{http://arxiv.org/abs/1203.5855}{{\tt arXiv:1203.5855}}.

\bibitem{Cheng:2013eu}
A.~Cheng, A.~Hasenfratz, G.~Petropoulos, and D.~Schaich
  \href{http://dx.doi.org/10.1007/JHEP07(2013)061}{{\em JHEP} {\bf 1307} (2013)
   061}, \href{http://arxiv.org/abs/1301.1355}{{\tt arXiv:1301.1355}}.

\bibitem{Aoki:2013xza}
LatKMI Collaboration, Y.~Aoki, T.~Aoyama, M.~Kurachi, T.~Maskawa, K.-i. Nagai,
  H.~Ohki, A.~Shibata, K.~Yamawaki, and T.~Yamazaki
  \href{http://dx.doi.org/10.1103/PhysRevD.87.094511}{{\em Phys. Rev.} {\bf
  D87} (2013)  094511}, \href{http://arxiv.org/abs/1302.6859}{{\tt
  arXiv:1302.6859}}.

\bibitem{Hasenfratz:2013uha}
A.~Hasenfratz, A.~Cheng, G.~Petropoulos, and D.~Schaich
  \href{http://arxiv.org/abs/1303.7129}{{\tt arXiv:1303.7129}}.

\bibitem{Aoki:2013qxa}
LatKMI Collaboration, Y.~Aoki, T.~Aoyama, M.~Kurachi, T.~Maskawa, K.~Miura,
  K.-i. Nagai, H.~Ohki, E.~Rinaldi, A.~Shibata, K.~Yamawaki, and T.~Yamazaki
  \href{http://arxiv.org/abs/1309.0711}{{\tt arXiv:1309.0711}}.

\bibitem{Hasenfratz:2013eka}
A.~Hasenfratz, A.~Cheng, G.~Petropoulos, and D.~Schaich {\em PoS} {\bf LATTICE
  2013} (2013)  075, \href{http://arxiv.org/abs/1310.1124}{{\tt
  arXiv:1310.1124}}.

\bibitem{Giusti:2008vb}
L.~Giusti and M.~Luscher
  \href{http://dx.doi.org/10.1088/1126-6708/2009/03/013}{{\em JHEP} {\bf 0903}
  (2009)  013}, \href{http://arxiv.org/abs/0812.3638}{{\tt arXiv:0812.3638}}.

\bibitem{Banks:1979yr}
T.~Banks and A.~Casher
  \href{http://dx.doi.org/10.1016/0550-3213(80)90255-2}{{\em Nucl. Phys.} {\bf
  B169} (1980)  103}.

\bibitem{DelDebbio:2010ze}
L.~Del~Debbio and R.~Zwicky
  \href{http://dx.doi.org/10.1103/PhysRevD.82.014502}{{\em Phys. Rev.} {\bf
  D82} (2010)  014502}, \href{http://arxiv.org/abs/1005.2371}{{\tt
  arXiv:1005.2371}}.

\bibitem{Patella:2012da}
A.~Patella \href{http://dx.doi.org/10.1103/PhysRevD.86.025006}{{\em Phys. Rev.}
  {\bf D86} (2012)  025006}, \href{http://arxiv.org/abs/1204.4432}{{\tt
  arXiv:1204.4432}}.

\bibitem{Landa-Marban:2013oia}
D.~Landa-Marban, W.~Bietenholz, and I.~Hip
  \href{http://arxiv.org/abs/1307.0231}{{\tt arXiv:1307.0231}}.

\end{thebibliography}\endgroup
}
\end{document}